# Gradient Optical Diffraction Tomography Using Deep Learning–Assisted Amplitude Estimation

**Julianna Winnik[1,*], Piotr Zdańkowski[1,*], Konrad Kosuge[1], Aleksandra Parys[1], Marzena Stefaniuk[2], Junwei Min[3], Baoli Yao[3,4], Maciej Trusiak[1]**

[1]Institute of Micromechanics and Photonics, Faculty of Mechatronics, Warsaw University of Technology, Warsaw, Poland
[2]Laboratory of Neurobiology, BRAINCITY, Nencki Institute of Experimental Biology of Polish Academy of Sciences, Warsaw, Poland
[3]State Key Laboratory of Ultrafast Optical Science and Technology, Xi'an Institute of Optics and Precision Mechanics, Chinese Academy of Sciences, Xi'an, 710119, China
[4]University of Chinese Academy of Sciences, Beijing, 100049, China
*Authors to whom any correspondence should be addressed.

E-mail: julianna.winnik@pw.edu.pl, piotr.zdankowski@pw.edu.pl

### Abstract

Gradient Optical Diffraction Tomography (GODT) enables the non-invasive reconstruction of a spatial derivative of the 3D sample's refractive index distribution, providing enhanced sensitivity to subtle refractive index variations through gradient-based contrast. Implemented using lateral shearing interferometry (LSI), GODT benefits from a common-path architecture that provides environmental stability and high signal-to-noise ratio measurements. Unlike conventional optical diffraction tomography (ODT), which requires numerical integration of LSI-derived phase gradients to recover the object phase, GODT enables direct tomographic reconstruction from the LSI data. Notably, GODT requires derivatives of both the phase and amplitude of the object fields as input. While the phase derivative can be readily extracted from LSI data, the amplitude gradient cannot be directly retrieved. Thus, in the original GODT formulation, the amplitude gradient was assumed to be negligible, limiting the accuracy of the resulting 3D reconstruction.

In this work, we introduce amplitude-informed GODT (AI-GODT), which employs physics-informed supervised deep learning to estimate the missing amplitude gradient based on accessible LSI data, specifically the fringe modulation. The proposed approach, applying a ResNet architecture and training on computer-generated data, is validated through numerical simulation and experimental measurements of both a well-characterized test object and primary neuronal cell culture, demonstrating the network's ability to generalize to diverse samples. We show that incorporation of the recovered amplitude gradient into the GODT reconstruction pipeline improves the 3D reconstruction quality, especially in its out-of-focus regions. The software and datasets are openly available to facilitate reproducibility and further development.

## 1. Introduction

Optical Diffraction Tomography (ODT) [1–3] is a label-free quantitative imaging technique that enables the non-invasive three-dimensional reconstruction of the refractive index (RI) distribution within transparent microscopic specimens, including biological cells [4] and tissue fragments [5]. By providing volumetric information on the optical properties of a sample, ODT has become a powerful tool for quantitative characterization of biological structures. However, as with other coherent phase-imaging techniques, conventional ODT is susceptible to coherent noise [6], which can substantially degrade image quality and limit the sensitivity to weak RI variations. Measurement robustness and the signal-to-noise ratio can be improved by using low-coherence illumination in a common-path interferometric configuration [7,8]. The common-path geometry provides intrinsic stability against environmental perturbations and maintains a small optical path difference between the superimposing beams, thereby enabling high interference-fringe contrast under low-coherence illumination while suppressing unwanted interference from multiply scattered [9,10] and parasitic [7] light. Importantly, it has been demonstrated [11,12] that measurements acquired under reduced light coherence can, under certain conditions, be processed using conventional reconstruction algorithms that assume coherent illumination.

There are various approaches to low-coherence ODT, each with its own advantages and disadvantages. Tomography using the diffraction phase microscopy principle [13,14] splits the object beam, and one of the resulting copies is spatially filtered in the Fourier plane to form a reference wave that is subsequently superimposed with the object beam. In another approach, the object beam is split into two copies that are sufficiently transversely separated such that one copy effectively interferes with an object-free region of the other [7,8]. Other spatially low-coherence ODT approaches adopt a fundamentally different strategy by replacing angular projections with $z$-stacking and 3D deconvolution. Among these approaches, spatial light interference tomography (SLIT) [9] applies a quantitative extension of conventional phase-contrast microscopy. Other methods combine $z$-stack acquisition with intensity-only techniques, such as structured illumination [15] and the weak object transfer function [16]. Low-coherence tomography can also be realized using a separate-arm Mach–Zehnder interferometric configuration, supplemented by a compensating diffraction grating in the reference arm and $z$-stack acquisition [17,18].

Another class of low-coherence common-path ODT methods [10,19–23] combines angular scanning with lateral shearing interferometry (LSI) [24–28]. These implementations offer simplicity, compactness, and hardware robustness. In these systems, the interference signal is generated by superimposing two slightly transversely displaced replicas of the object wave, enabling interferometric phase coding under low-coherence conditions. However, rather than providing the object phase directly, LSI measures the phase difference across the imposed lateral shear, which approximates the spatial derivative of the object phase. This derivative-based measurement is advantageous for robust interferometric detection but introduces additional requirements for tomographic reconstruction. In conventional processing pipelines [10,19–22], the measured shearing phase maps must first be numerically integrated to recover the corresponding object phase distributions before applying a standard ODT reconstruction algorithm. This additional processing step can be computationally demanding and is sensitive to measurement noise, boundary conditions, and integration artifacts, potentially introducing errors into the reconstructed 3D RI distribution. Moreover, recovering the absolute phase map required for tomographic reconstruction generally requires measurements for two orthogonal shear directions. This necessitates either sequential acquisition with different shear orientations [27] or additional optical complexity [29] and a reduced space–bandwidth product [25], increasing the acquisition burden and potentially introducing inconsistencies between measurements.

Recent work [23] introduced Gradient Optical Diffraction Tomography (GODT), a 3D reconstruction framework that enables direct recovery of the spatial derivative of the three-dimensional refractive-index (RI) distribution from a tomographic sequence of shearing interferometry measurements. By directly reconstructing the RI derivative, GODT eliminates the need for numerical phase integration, thereby simplifying the processing pipeline and avoiding an important source of error. Additionally, GODT mitigates phase-wrapping issues [30] and enhances the visualization of subtle RI variations [23].

GODT is derived under the first-order Rytov approximation [31] and, like conventional ODT approaches such as direct inversion and filtered backpropagation [32], requires complex-valued information about the object field, comprising both phase and amplitude parts. While the phase derivative can be readily obtained from LSI interferograms, the corresponding amplitude derivative is not directly accessible. In the original implementation of GODT [23], this problem was addressed by assuming a uniform amplitude of the object waves and, consequently, a zero amplitude derivative. This approximation enables GODT to operate using phase data alone and yield accurate results for thin, weakly scattering transparent objects. However, it can compromise reconstruction accuracy when the object field exhibits significant amplitude variations. This limitation is particularly pronounced in out-of-focus regions of the 3D object, where defocusing induces substantial spatial variations in the field's amplitude. Neglecting these variations can introduce reconstruction artifacts and reduce the fidelity of the recovered RI gradient distribution.

In this work, we significantly enhance GODT by incorporating physics-informed supervised deep learning to overcome its fundamental limitation: the lack of amplitude-derivative information. The proposed pipeline exploits the modulation of the LSI fringe patterns, which encodes the product of the amplitudes of two laterally displaced copies of the object field. This experimentally accessible signal is used as input to a convolutional ResNet-based [33] neural network trained on computer-generated data to estimate the corresponding amplitude input, i.e. normalized amplitude derivative, required for GODT. We validate the proposed amplitude-informed GODT (AI-GODT) using both numerical simulation and experimental measurements of a well-characterized test object and neurons in vitro. The results demonstrate that the proposed deep network–based pipeline generalizes from simulated to experimental data, maintaining its performance under variations in system parameters and object characteristics. The AI-GODT results demonstrate superior reconstruction quality compared with the conventional phase-only GODT, particularly in out-of-focus regions.

To facilitate further development and adoption of the method, the source code [34] and test data [35] are made openly available.

## 2. Amplitude-informed GODT

### 2.1 Motivation

ODT reconstructs 3D refractive index distribution $n(x, y, z)$ based on a stack of object fields $u_\theta(x, y)$ acquired subsequently for a range of illumination plane waves $u_\theta^{\text{ill}}$ propagating at angles $\theta$. Using the first-order Rytov approximation of the generalized projection theorem, each field $u_\theta$ is first logarithmically transformed:

$$\ln[u_\theta(x,y)] = \ln[A_\theta(x,y)] + j\varphi_\theta(x,y), \tag{1}$$

where $A_\theta$ and $\varphi_\theta$ denote amplitude and phase of the normalized object field $u_\theta/u_\theta^{ill}$. The stack of Rytov fields is subsequently processed by the ODT reconstruction algorithm, which maps the spatial-frequency information contained in each field onto the corresponding Ewald sphere within the 3D sample spectrum [31].

GODT algorithm is very similar to ODT, with the difference that it operates on the spatial derivative of the conventional ODT input (Eq. (1)), which yields:

$$\frac{\partial \ln[u_\theta]}{\partial s} = \frac{1}{A_\theta}\frac{\partial A_\theta}{\partial s} + j\frac{\partial \varphi_\theta}{\partial s}, \tag{2}$$

where $\boldsymbol{s}$ is the small shear vector lying in the $(x, y)$ plane:

$$\boldsymbol{s} = (cos\gamma, sin\gamma, 0). \tag{3}$$

Thereby, GODT reconstructs the corresponding spatial derivative of the ODT output, namely, the refractive index distribution differentiated along $\boldsymbol{s}$ [23]:

$$\frac{\partial n}{\partial s} = \boldsymbol{s} \cdot \nabla n. \tag{4}$$

Equation (2) defines the required input for GODT. Its imaginary part $\frac{\partial \varphi_\theta}{\partial s}$, can be readily obtained using LSI, in which the interferogram is generated by superposing two copies of the object field displaced by a small shear $\boldsymbol{s}$:

$$I(\boldsymbol{r}) = |u_\theta(\boldsymbol{r}) + u_\theta(\boldsymbol{r}+\boldsymbol{s})|^2, \tag{5}$$

which can be expanded as:

$$I(\boldsymbol{r}) = {A_\theta}^2(\boldsymbol{r}) + {A_\theta}^2(\boldsymbol{r}+\boldsymbol{s}) + 2A_\theta(\boldsymbol{r})A_\theta(\boldsymbol{r}+\boldsymbol{s})cos[\varphi_\theta(\boldsymbol{r}) - \varphi_\theta(\boldsymbol{r}+\boldsymbol{s})]. \tag{6}$$

The argument of the cosine in Eq. (6) is directly related to the spatial derivative of the object phase along the shear direction:

$$\varphi_\theta(\boldsymbol{r}) - \varphi_\theta(\boldsymbol{r}+\boldsymbol{s}) \approx s\frac{\partial \varphi_\theta}{\partial s}. \tag{7}$$

The problem, however, arises with the required amplitude input for GODT, i.e., the real part of Eq. (2), which is the normalized derivative of the object amplitude: $\frac{1}{A_\theta}\frac{\partial A_\theta}{\partial s}$. This distribution cannot be directly extracted from the LSI fringe pattern. Nevertheless, the amplitude information is encoded in the fringe modulation, i.e., $2A_\theta(\boldsymbol{r})A_\theta(\boldsymbol{r}+\boldsymbol{s})$. AI-GODT takes advantage of this information content by applying a supervised deep-learning approach to infer the required amplitude input for GODT from the accessible amplitude-related distribution:

$$A_\theta(\boldsymbol{r}) \cdot A_\theta(\boldsymbol{r}+\boldsymbol{s}) \quad \xrightarrow{DNN} \quad \frac{A_\theta(\boldsymbol{r}) - A_\theta(\boldsymbol{r}+\boldsymbol{s})}{A_\theta(\boldsymbol{r})}. \tag{8}$$

Notably, the discussion above, particularly Eqs. (2) and (4), assume an infinitesimal shear. However, a larger shear does not compromise the accuracy of the GODT algorithm [23]. For an arbitrary finite shear magnitude, the GODT inputs can be expressed as finite differences:

$$\frac{1}{|s|}\left\{\frac{A_\theta(\boldsymbol{r}) - A_\theta(\boldsymbol{r}+\boldsymbol{s})}{A_\theta(\boldsymbol{r})} + j[\varphi_\theta(\boldsymbol{r}) - \varphi_\theta(\boldsymbol{r}+\boldsymbol{s})]\right\}, \tag{9}$$

and the corresponding GODT output is likewise a finite-difference approximation of the spatial derivative of the refractive index distribution:

$$\frac{1}{|s|}\{n(\boldsymbol{r}, z) - n(\boldsymbol{r}+\boldsymbol{s}, z)\}. \tag{10}$$

### 2.2 Pipeline

The full pipeline of AI-GODT is presented in Fig. 1(a). The sequence of low-coherence LSI interferograms corresponding to various illumination directions is reconstructed to obtain a set of low-noise fringe modulations and phases. The fringe phases can be directly used as the input to GODT. In contrast, the fringe modulations require transformation by a deep neural network (DNN) to obtain the required amplitude input for GODT. Finally, GODT retrieves the spatial derivatives of the refractive index distribution. The GODT tomographic evaluation is performed using open-source code [36], based on an analogous ODT algorithm [37].

The applied DNN training strategy is shown in Fig. 1(b). To learn the required mapping between the fringe modulation and the normalized amplitude derivative, we employ supervised training of a lightweight convolutional ResNet model. Convolutional layers extract spatially localized features and hierarchical representations of fringe modulation patterns, enabling the network to learn their relationship with the normalized amplitude derivative, while residual connections facilitate gradient propagation and mitigate the vanishing-gradient problem. The DNN model is trained once for a given shear $\boldsymbol{s}$ and can then process all datasets acquired with the same $\boldsymbol{s}$. The training is achieved using computer-generated training data constructed based on the physical relationship between the fringe modulation and the required GODT input, using natural images obtained from an open-source database as source distributions for the object field amplitude. This strategy promotes good generalization by providing a large diversity of morphological features, similarly to [38,39]. Each natural image is used to generate a training sample pair consisting of (i) a product of sheared images that emulates the LSI fringe modulation and (ii) the corresponding desired input for GODT, namely, the normalized image derivative. The former serves as the input, while the latter serves as the target during supervised training.

The proposed deep-learning approach is computationally efficient and practical, as it relies on synthetically generated training data and can be trained in approximately 15 minutes on a laptop equipped with an NVIDIA RTX 4080 GPU. These modest data and computational requirements facilitate the practical implementation of AI-GODT without the need for extensive experimental training datasets or specialized high-performance computing infrastructure. Details of data generation, model architecture, training procedure, and model evaluation are provided in Appendices A, B, C, and D, respectively.

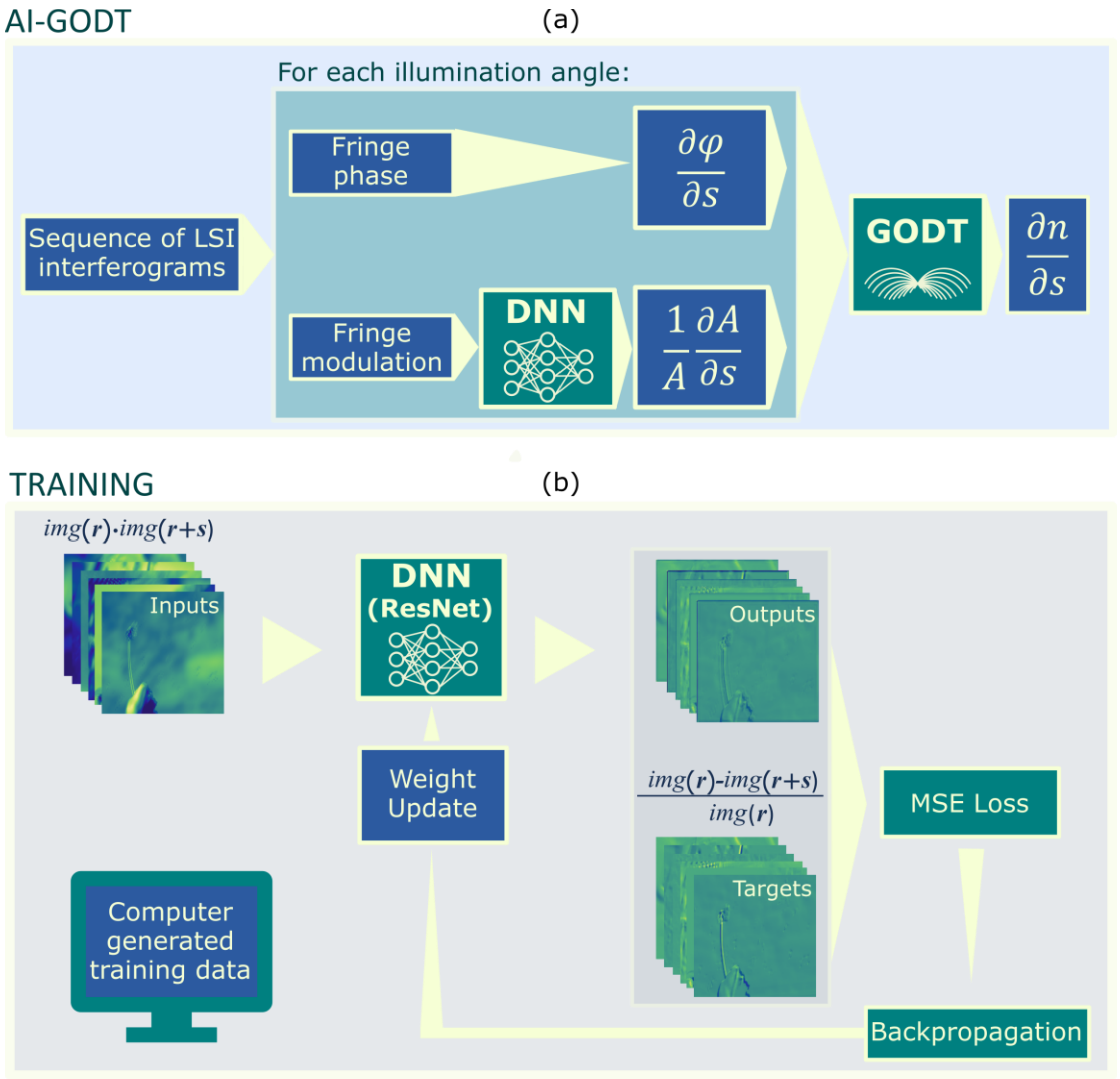

**Figure 1.** Schematic illustration of the AI-GODT pipeline (a) and the deep neural network (DNN) training procedure (b).

## 3. Results

### *3.1 Simulation study*

In this section, we first evaluate the impact of neglecting amplitude information in GODT and then assess the performance of the proposed AI-GODT in addressing this issue. We use a modified Shepp–Logan phantom (Fig. 2(a)) with a refractive index ranging from 1.334 to 1.340, immersed in a medium with refractive index $n_0$ = 1.330. We simulate 180 object waves under conical illumination scenario, with a 60° inclination relative to the optical axis and a 2° azimuthal step. The simulation uses λ = 0.5 μm and numerical aperture of the optical system NA = 1.33. Wave propagation through the inhomogeneous 3D object is simulated using the multi-slice wave propagation method (WPM) [40,41], with 1000 × 1000-pixel object waves sampled at 0.1 μm.

The simulated object fields are then processed to obtain the LSI data required for this test: the ideal GODT input (Eq. (9)) and the fringe modulation used in AI-GODT (the left-hand side of Eq. (8)). We set the shear to $\boldsymbol{s}$ = [6.6, -8] pixels, consistent with the value used in the subsequent experimental evaluation. We then perform GODT reconstruction under three scenarios: (i) using the ideal input; (ii) using phase data alone, with the amplitude information neglected by assuming a zero-amplitude derivative; and (iii) using AI-GODT, in which the required amplitude input is estimated from the simulated fringe modulation by the previously trained DNN. The reconstruction results are presented in Fig. 2(b)–(c).

Figure 2(c) shows three transverse and one axial cross-section of the 3D reconstructions obtained using reconstruction strategies (i)–(iii), with the section locations indicated in Fig. 2(a). A through-focus video showing the transverse cross-sections sequentially is also provided (Visualization 1). Figures 2(c1)–(c4) show the reference GODT reconstruction obtained using the ideal input, which enables clear visualization of the phantom's refractive-index gradient. This result is used as ground truth for our test. Figures 2(c5)–(c8) illustrate the effect of omitting the amplitude input. The resulting reconstruction exhibits blurring, related contrast loss as well as diffraction-like artefacts. These effects become increasingly pronounced away from the focal plane ($z = 0$ denotes the sample plane that is optically conjugated with the camera). This behaviour arises because amplitude variations are primarily induced by defocusing and become more pronounced with increasing distance from the focal plane, making amplitude information increasingly important for accurate reconstruction. Without the information-bearing amplitude derivatives, GODT has limited ability to invert diffraction effects, leading to reduced tomographic reconstruction fidelity in out-of-focus planes. This trend is clearly reflected in Fig. 2(b), where the blue curve shows the normalized root-mean-square error:

$$\mathrm{NRMSE}(z) = \frac{\left\| T(z) - T_{ref}(z) \right\|_2}{\left\| T_{ref}(z) \right\|_2} \tag{11}$$

of the subsequent transverse sections of the tomographic reconstruction $T$ relative to the reference ideal-input GODT reconstruction $T_{ref}$.

Next, we evaluate the performance of the proposed AI-GODT reconstruction strategy. As shown in Figs. 2(c9)–(c12), AI-GODT substantially reduces the structural blurring and diffraction-like artifacts observed when amplitude information is neglected (Figs. 2(c5)–(c8)). This observation is also quantitatively confirmed by the NRMSE$(z)$ plot shown in Fig. 2(b), where the red curve represents AI-GODT and the blue curve represents GODT without amplitude information. Over the defocused region ($|z|$ > 3 μm), incorporating amplitude information via AI-GODT reduces the reconstruction error by an average of 54%.

Regarding computational requirements, training the DNN model took approximately 15 minutes. The inference time for 180 normalized amplitude derivatives was approx. 30 seconds. The computational time required for GODT was the same for phase-only GODT and AI-GODT, amounting to approximately 55 seconds for the considered case.

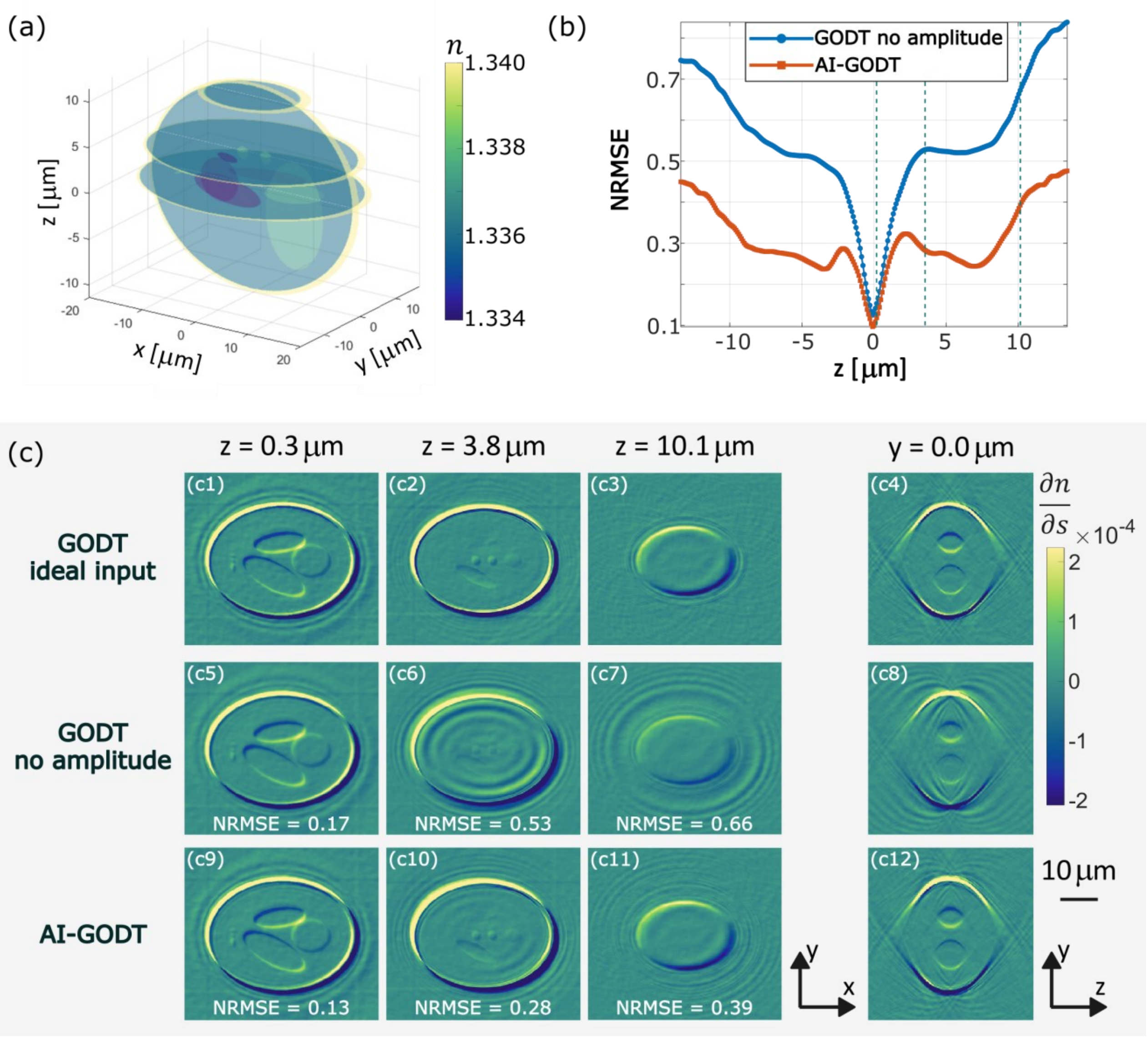

**Figure 2.** Simulation test: (a) 3D view of the phantom; (b) normalized root-mean-square error (NRMSE) of the transverse sections of the 3D reconstructions as a function of an axial distance. The ground truth used for the NRMSE calculation is the GODT reconstruction obtained with the ideal input; *z=0* corresponds to the in-focus plane. The blue and red curves represent GODT without amplitude information and AI-GODT, respectively; (c) representative transverse and axial cross-sections of the 3D reconstructions obtained using (c1)–(c4) GODT with the ideal input, (c5)–(c8) GODT without amplitude information, and (c9)–(c12) AI-GODT.

### *3.2 Experimental study*

In this work, the AI-GODT method is experimentally demonstrated with the common-path LSI system based on liquid-crystal polarization gratings [8,23,27] shown in Fig. 3. A red LED (Thorlabs M625L4; λ = 625 nm, Δλ = 17 nm FWHM) illuminates the sample through a collimating lens and a linear polarizer, $P_{LED}$. The finite spectral bandwidth provides partial temporal coherence leading to high signal-to-noise-ratio phase measurements. An automated rotary arm scans uniformly distributed illumination directions (usually 180) at a fixed illumination tilt angle $\alpha$, thereby tracing a cone around the optical axis. The system works in the inverted microscope geometry, where the microscope objective (MO) is placed below the sample and together with the tube lens (TL) forms the imaging system with transverse magnification of $M$. The lateral-shearing module is located between the TL and the camera CMOS. It consists of two identical, parallel polarization diffraction gratings, $PG_1$ and $PG_2$, separated by a small axial distance $d_{PG}$. $PG_1$ splits the linearly polarized object wave into the +1 and −1 diffraction orders, which propagate at angles $+\theta_d$ and $-\theta_d$ with opposite circular polarizations. $PG_2$ reverses their handedness and redirects the two copies into parallel propagation. The resulting transverse displacement $\Delta'$ between the object beams is given by

$$\Delta' = 2 d_{PG} \tan\theta_d. \tag{12}$$

The corresponding object-space shear magnitude, $\Delta$, is obtained by scaling by the inverse transverse magnification 1/M, i.e. $\Delta = \Delta' / M$.

The two copies have opposite circular polarizations and must therefore be projected onto a common linear polarization to interfere. Because the relative phase shift between the beams depends on the orientation of the polarizer's transmission axis, this projection also enables controlled phase shifting (PS) of the fringe patterns, thereby facilitating phase demodulation. For the phase shifting we used two detection configurations (Fig. 3(c)). The first configuration uses a polarization-resolved camera with a micro-polarizer array that simultaneously samples four analyzer orientations across interleaved pixel subsets [27]. The second detection configuration performs temporal phase shifting (TPS) [8] using a rotating linear analyzer, $P_{CAM}$, followed by a standard CMOS camera (Fig. 3(a)). Rotating the analyzer through $\beta$ changes the relative phase of the projected components by $2\beta$. PS-based interferogram demodulation yields the fringe phase and modulation, which in LSI correspond to the object phase difference across the imposed shear and the product of the sheared object-field amplitudes (Eq. (6)), respectively.

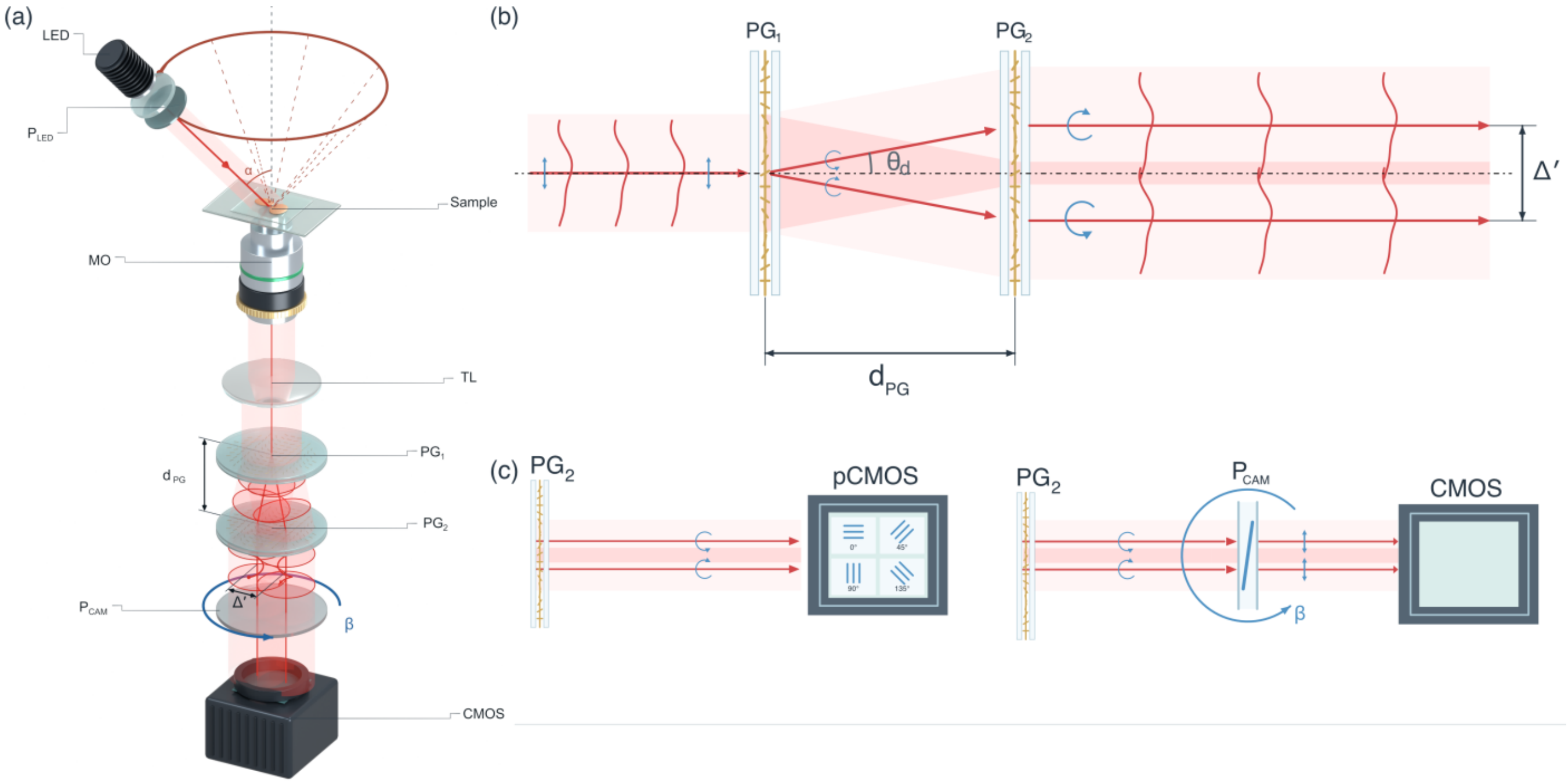

**Figure 3.** Polarization diffraction grating-based LSI system. (a) Schematic of the overall system: LED – light emitting diode; $P_{LED}$ – linear polarizer; MO – microscope objective; TL – tube lens; $PG_1$ and $PG_2$ – liquid-crystal polarization diffraction gratings; $P_{CAM}$ – linear analyzer. (b) Schematic of the shearing module: $PG_1$ generates two diffraction orders at angles $\pm\theta_d$ with opposite circular polarizations, while $PG_2$ reverses their handedness and restores parallel propagation; $d_{PG}$ is the axial gratings separation and $\Delta'$ is the transverse displacement between the object wave copies. The copies interfere after projection onto a common linear polarization. (c) Alternative detection schemes left – a polarization-resolved camera for single-shot phase shifting and right – a CMOS camera for temporal phase shifting achieved by rotating the linear polarizer $P_{CAM}$.

We first evaluate AI-GODT using a well-defined object—a phase target in the shape of the letter G, manufactured by Lyncée Tec by etching 125 ± 5 nm-deep structures into Borofloat 33 glass. The object was deliberately defocused during the measurement to assess the accuracy of the tomographic reconstruction in the out-of-focus region. For this experiment, we used a microscope objective with numerical aperture NA = 0.75 and magnification $M = 20$. The illumination tilt angle was set to $\alpha = 33.5°$, and the rotary arm scanned 72 uniformly distributed illumination directions. PS was enabled using a polarization-resolved camera (pixel pitch of 3.45 $\mu$m) incorporating a micro-polarizer array (Fig. 3(c)). The system was set to produce a small shear, i.e., $\boldsymbol{s} = [-1.11, 0.80]$ $\mu$m. The tomographic series of LSI interferograms was demodulated using PS, yielding the fringe phase and contrast. The resulting data were then upsampled to facilitate the tomographic reconstruction, resulting in a final sampling pitch of 0.26 $\mu$m. Using this sampling the shear vector yields $\boldsymbol{s} = [-4.3, 3.1]$ pixels. The processed data were subsequently fed into the AI-GODT pipeline shown in Fig. 1.

The tomographic evaluation of the phase target is presented in Fig. 4. Figures 4(a) and 4(b) show transverse and axial cross-sections, respectively, of the 3D reconstruction obtained using standard GODT with the amplitude input omitted. The axial cross-section in Fig. 4(b) reveals the target positioned 10 μm away from the in-focus plane. The reconstructed structure exhibits noticeable blurring. In contrast, the corresponding cross-sections obtained using the proposed AI-GODT approach (Figs. 4(c) and 4(d)) exhibit improved sharpness. This enhancement is further illustrated by enlarged views of the regions marked in Figs. 4(a) and 4(c), presented in Figs. 4(e) and 4(f), respectively, and by the 1D profiles in Fig. 4(g), which demonstrate increased contrast of the AI-GODT result. Quantitative evaluation using the full width at half maximum (FWHM) of the positive and negative lobes of the tomographic reconstruction confirms this improvement (Fig. 4(h)). The respective widths decrease from 2.93 to 2.07 µm and from 2.65 to 1.99 µm. Additionally, a through-focus comparison of the standard GODT and AI-GODT reconstructions is provided in Visualization 2.

These results demonstrate that the DNN, although trained on synthetically generated data, generalizes to experimental measurements and effectively estimates the amplitude input required for GODT from the fringe modulation. The proposed approach therefore enables amplitude-informed GODT reconstruction even when direct amplitude information is not experimentally available.

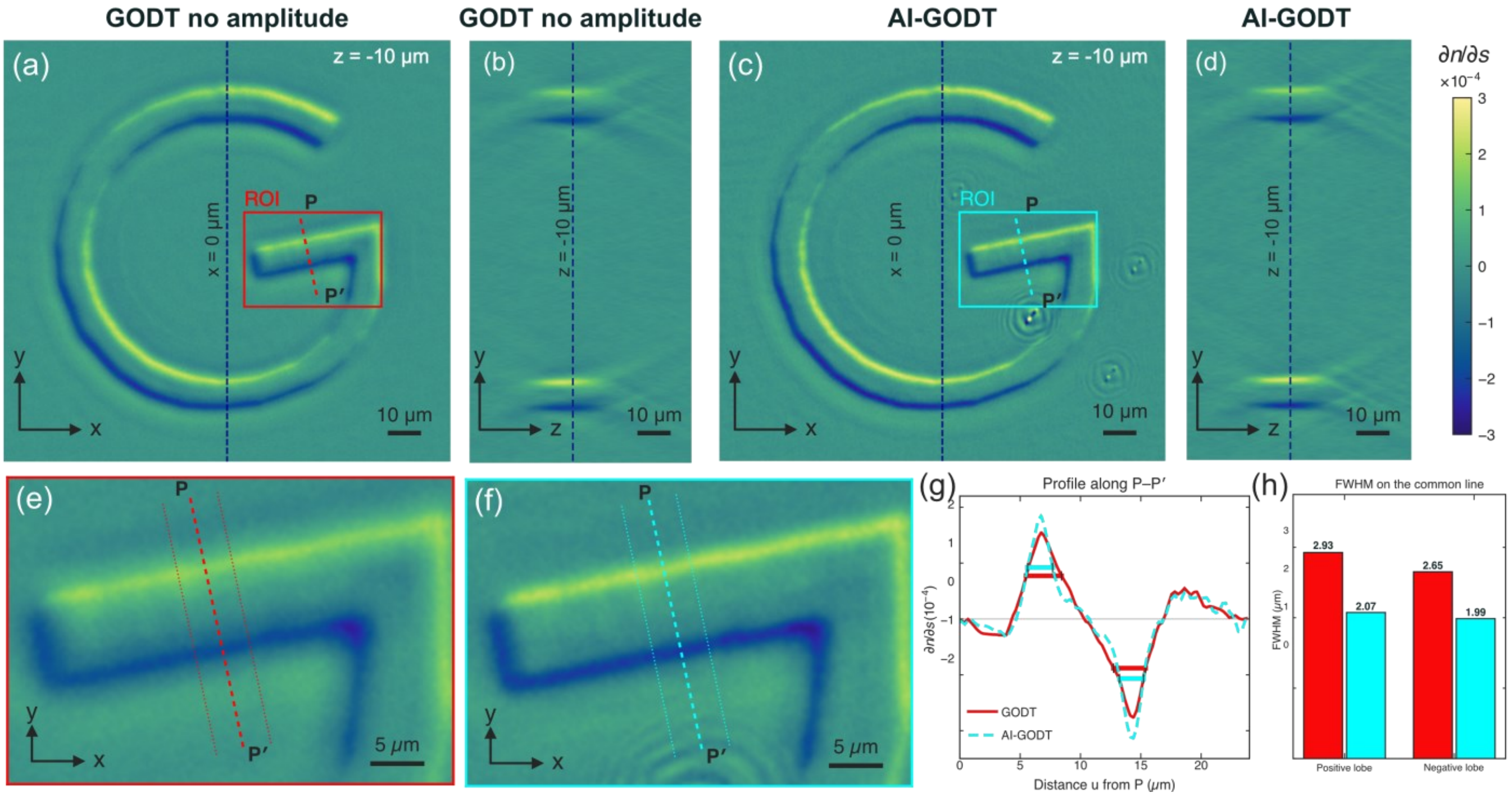

**Figure 4.** Experimental tomographic reconstruction of a phase target using standard GODT and the proposed AI-GODT approach. (a, b) Transverse and axial cross-sections, respectively, reconstructed using standard GODT with the amplitude input omitted. (c, d) Corresponding cross-sections obtained using AI-GODT. (e, f) Enlarged views of the regions indicated in (a) and (c), respectively. (g) Comparison of the 1D reconstruction profiles and (h) full width at half maximum (FWHM) of the positive and negative lobes, demonstrating improved structural contrast and spatial resolution provided with AI-GODT. The through-focus comparison of both methods is provided in Visualization 2.

Next, we used AI-GODT for measurement of hippocampal neurons obtained from newborn rats. In this case, we used an oil-immersion objective with a numerical aperture of NA = 1.45 and a magnification of $M$ = 100. The illumination tilt angle was set to $\alpha$ = 70°, and 180 uniformly distributed illumination directions were scanned using the rotary arm. Phase shifting was realized using a rotary polarizer placed at the end of the LSI setup (Fig. 3(c)). The data was captured using a camera with a pixel pitch of 3.45 µm. The shear vector was $\boldsymbol{s}$ = [0.86, -1.04] µm, which corresponds to [6.6, -8] pixels (after resampling of data to final sampling pitch of 0.13 µm).

The tomographic reconstruction results are presented in Figs. 5 and 6, showing four selected transverse cross-sections at different depths for two neuronal cell samples. In each case, we compare GODT without amplitude information (a)–(d) with AI-GODT (e)–(h). Through-stack videos are also provided: Visualization 3 for the sample in Fig. 5 and Visualization 4 for that in Fig. 6. The experimental comparison of the phase-only GODT and AI-GODT is consistent with the simulation results presented in Fig. 2(c). Both approaches produce comparable reconstruction quality near the in-focus plane, as shown in Figs. 5(b), (f) and 6(c), (g). However, away from the in-focus plane, the omission of amplitude information leads to image blurring and a corresponding loss of contrast, as observed in Figs. 5(a), (c), (d) and 6(a), (b), (d). This degradation is substantially reduced by AI-GODT, as demonstrated in Figs. 5(e), (g), (h) and 6(e), (f), (h).

The improved reconstruction quality achieved by AI-GODT enables better visualisation of fine neuronal structures across different depths. In particular, Figs. 5(e) and 6(e), (f) demonstrate significantly enhanced visibility of neuronal projections and the extracellular matrix (ECM) surrounding neuronal cells. The increased spatial resolution and contrast reveal fine structural details that are difficult to distinguish using the conventional GODT approach. This facilitates a more precise characterisation of ECM organisation, including the distribution, morphology and spatial relationships of matrix components in close proximity to neuronal cell bodies and processes.

The improved resolution also facilitates the identification of subtle cellular morphological features, including those associated with apoptosis. In particular, the neuron shown exhibits pronounced apoptotic fragmentation, characterised by the formation of distinct, rounded apoptotic bodies resulting from the disintegration of the cell body and its components. These spherical structures are indicative of the progressive dismantling of the neuron into membrane-bound apoptotic bodies. Compared with the conventional approach, AI-GODT provides improved clarity, contrast and spatial resolution, allowing these small-scale morphological features to be distinguished more clearly across different depths.

The advantages of AI-GODT become particularly evident at larger distances from the in-focus plane, where stronger defocusing effects are present in the captured data. In these regions, the absence of amplitude information limits the ability of the tomographic reconstruction algorithm to effectively invert diffraction effects, resulting in diffraction-like artefacts. An example of such artefacts is visible in the region indicated by the arrows in Fig. 5(d). These artefacts are substantially suppressed by AI-GODT, as shown in Fig. 5(h).

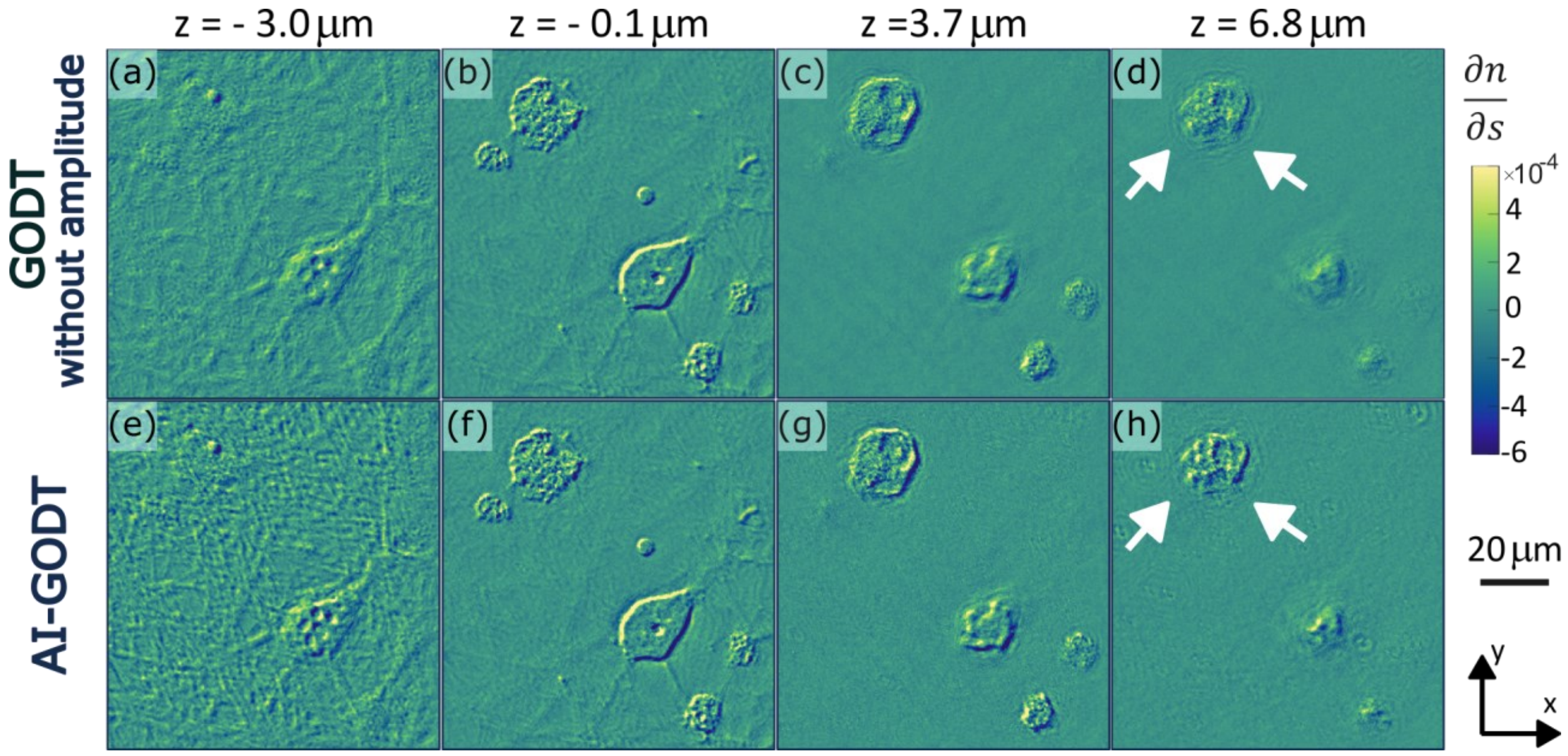

**Figure 5.** Tomographic reconstruction of the neuronal cells sample no. 1. (a)–(d) GODT without amplitude information and (e)–(h) AI-GODT, shown at four selected depths. The corresponding through-stack video is provided in Visualization 3.

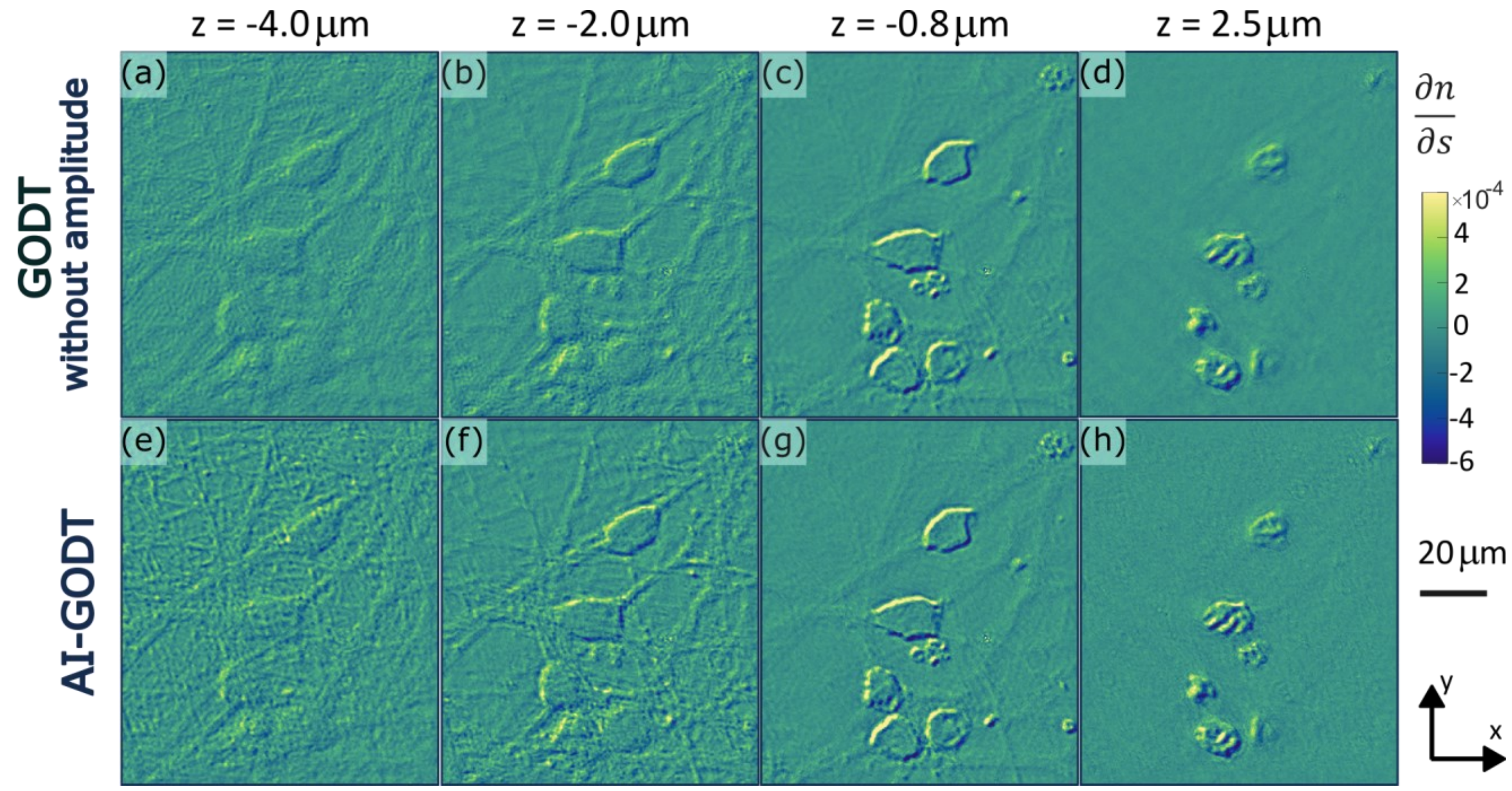

**Figure 6.** Tomographic reconstruction of the neuronal cells sample no. 2. (a)–(d) GODT without amplitude information and (e)–(h) AI-GODT, shown at four selected depths. The corresponding through-stack video is provided in Visualization 4.

## Discussion

The results demonstrate the importance of incorporating amplitude information into GODT to improve the quality of 3D reconstructions. The phase-only GODT, introduced in [23], offers a relatively simple reconstruction framework, requiring only phase information from the measured fringe patterns, and is generally sufficient for thin, transparent objects, for which amplitude variations are negligible. However, increased sample thickness, scattering, or absorption can induce significant amplitude variations that carry additional information about the object, the neglect of which may compromise 3D reconstruction accuracy. Among the various sources of amplitude variations in optical tomography, sample defocusing is of particular relevance and is therefore the focus of this work. Defocus-introduced amplitude variations become increasingly pronounced with the distance of a sample feature from the focal plane. Consequently, phase-only GODT exhibits reduced reconstruction accuracy in the out-of-focus planes of the reconstructed volume, which may limit its effectiveness for thick samples. This loss of reconstruction fidelity in the out-of-focus region is demonstrated through numerical simulations (Fig. 2(b)) and experimental measurements using a well-characterized test target (Fig. 4).

The proposed AI-GODT approach enhances the original phase-only GODT by overcoming its inherent limitation through the recovery of missing amplitude information from fringe modulation using supervised deep learning. This improves reconstruction fidelity, particularly in out-of-focus regions, without compromising the key advantages of GODT, namely its high sensitivity to small refractive index variations and its compatibility with low-coherence, common-path LSI systems that enable high-SNR phase-derivative measurements. Importantly, deep learning is used solely to estimate the additional amplitude input, while the tomographic reconstruction itself relies on the analytical GODT solver. This preserves the computational efficiency of the original approach, enabling the reconstruction of a $1000 \times 1000 \times 1000$ voxel volume in approximately one minute.

The improved reconstruction quality achieved by AI-GODT compared with phase-only GODT, as demonstrated with numerical simulations (Fig. 2) and experiments (Fig. 4), confirms the effectiveness of the proposed strategy. In simulations, AI-GODT reduced the reconstruction error by an average of 54% in the defocused region. The estimated amplitude derivatives provide additional information for diffraction compensation, improving structural contrast and reconstruction fidelity, particularly for features farther from the focal plane, where defocus-induced amplitude variations are more pronounced (Fig. 2(b)).

The experiments on neuronal cells (Figs. 5 and 6) further demonstrate the potential of AI-GODT for label-free biological imaging. The improved visibility of fine neuronal structures, including neuronal projections, the extracellular matrix (ECM) surrounding neuronal cells, and subtle cellular morphological features, in the reconstructed refractive index gradient highlights the ability of the proposed approach to reveal structural details that are difficult to distinguish using conventional phase-only GODT. In particular, the enhanced visualisation of cellular morphology, including features associated with apoptotic fragmentation, demonstrates the potential of AI-GODT for resolving fine structural details across different imaging depths.

Such capabilities are particularly relevant for applications in which preserving intrinsic cellular properties and avoiding exogenous contrast agents are important.

Although demonstrated using a polarization diffraction grating-based LSI system [8,23,27], AI-GODT is not limited to this configuration and can be adapted to other LSI implementations. An alternative approach to obtaining the required normalized amplitude gradient is to calculate it numerically from separately acquired amplitude measurements of the two laterally sheared object beams. However, this strategy presents practical limitations. First, isolating and measuring the individual beam amplitudes requires blocking one of the interfering beams, which is not always feasible in common-path LSI configurations. Second, tomographic imaging typically requires the acquisition of numerous angular projections, ranging from several dozen to several hundred, making separate amplitude measurements for each projection experimentally cumbersome. In this context, estimating the amplitude information directly from fringe modulation offers an attractive alternative, as it avoids the need for additional acquisitions and can be integrated into the existing LSI measurement workflow.

An important aspect of AI-GODT is the ability of the DNN model to generalize from synthetically generated training data to experimental measurements. This generalization is facilitated by the fact that the DNN is trained to learn the underlying physics-based mathematical relationship between LSI fringe modulation and the required GODT input, rather than simply memorizing the training samples. Furthermore, the generalization capability is supported by the broad diversity of morphological features extracted from natural images, which are used to generate the object amplitude distributions without introducing bias toward any particular sample type. This strategy eliminates the need for extensive experimentally acquired training datasets, which are often difficult to obtain and may lead to over-specialization to specific experimental configurations or sample types. Moreover, network training takes approximately 15 minutes on a mid-range scientific computer, demonstrating that the approach does not require specialized high-performance computing infrastructure. Although the network must be trained separately for each shear vector, the short training time makes this requirement readily manageable.

AI-GODT shares similar limitations with conventional GODT, as both approaches rely on the first-order Rytov approximation and therefore do not account for multiple scattering. Additionally, the method requires prior knowledge of the shear vector, which may be difficult to determine experimentally. Moreover, AI-GODT leads to a slightly increased noise level in the reconstructed results, as can be observed in the object-free regions of the reconstructed neuronal cells (e.g., by comparing Fig. 6(h) with Fig. 6(d)). This is because GODT, under the uniform amplitude assumption, avoids noise contributions arising from amplitude data. Furthermore, the neural network used in AI-GODT performs a task analogous to differentiating the amplitude distribution, which may amplify noise present in the experimentally acquired data.

Future work will explore alternative network architectures to further improve the accuracy of normalized amplitude derivative estimation. We also plan to investigate, through numerical simulations and experiments, additional mechanisms of amplitude variation in the object field, including absorptive properties of the sample.

## Conclusions

In this work, we introduced AI-GODT, an extension of gradient optical diffraction tomography (GODT) that incorporates amplitude derivative information estimated from fringe modulation using supervised learning. The proposed approach was evaluated through numerical simulations and experiments involving a well-characterized phase target and primary hippocampal cell culture. The results demonstrate that incorporating estimated amplitude information improves reconstruction fidelity and structural contrast compared with conventional phase-only GODT and the difference is especially evident in out-of-focus regions of the 3D reconstruction. Experiments on neuronal cells further demonstrate the potential of AI-GODT for label-free visualisation of fine cellular structures and neuronal morphology at different imaging depths through refractive index gradient reconstruction.

Notably, the proposed network successfully generalized from synthetic training data to experimental measurements. Moreover, it required only approximately 15 minutes of training on a mid-range scientific computer. Together, these results highlight the practical advantages of AI-GODT: it requires no experimentally acquired training data and can be trained quickly using modest computational resources. Its computational efficiency is further enhanced by the proposed hybrid approach, in which the DNN is used solely to estimate the missing amplitude information, while the final tomographic reconstruction is performed using an efficient, analytical, physics-based solver. This separation also helps preserve the reliability and physical consistency of the reconstruction results.

Overall, AI-GODT demonstrates the potential of combining synthetic-data-based learning with physics-based reconstruction, in which deep learning complements rather than replaces the underlying diffraction tomography framework. By estimating amplitude information missing from conventional phase-only GODT, this hybrid approach offers a practical means of improving the reliability and applicability of derivative-based optical tomography in experimental imaging.

The implementation is made openly available to support reproducibility and facilitate further development of machine-learning-assisted optical tomography.

## CRediT authorship contribution statement

**Julianna Winnik**: Methodology, Investigation, Data curation, Writing – original draft, Visualization, Software.

**Piotr Zdańkowski**: Conceptualization, Methodology, Investigation, Writing – original draft, Visualization, Funding acquisition, Supervision, Project administration.

**Konrad Kosuge**: Writing – original draft, Software, Visualization.

**Aleksandra Parys**: Investigation.

**Marzena Stefaniuk**: Resources, Writing – original draft.

**Junwei Min**: Writing – review & editing.

**Baoli Yao**: Funding acquisition, Project administration, Writing – review & editing.

**Maciej Trusiak**: Supervision, Funding acquisition, Project administration, Methodology, Conceptualization, Writing – review & editing.

## Funding

This work was supported by the National Science Center, Poland (grant no. 2024/55/D/ST7/02792); project no. WPC3/2022/47/INTENCITY/2024, funded by the National Center for Research and Development (NCBR) under the 3rd Polish-Chinese/Chinese-Polish Joint Research Call (2022); and the CAS President's International Fellowship Initiative (PIFI) group program (grant no. 115HZJL20250012P). MS was supported by the Foundation for Polish Science under the European Funds for Smart Economy project BRAINCITY (FENG.02.01-IP.05-0010/24), co-financed by the European Union.

## Acknowledgement

We thank Matyáš Heto for his assistance with the registration of tomographic projections of the test object.

## Data availability

The test data used in this study will be made publicly available upon publication at the repository cited in [35].

## Software availability

The software developed to estimate the missing amplitude information in AI-GODT will be made publicly available upon publication at the repository cited in [34]. The GODT reconstruction algorithm is available in [36].

## Appendix A. Training data

The training dataset was generated numerically using natural images as the source distributions to define the spatial distributions of the object-wave amplitude. This approach promotes diversity in the training data and thus reduces the risk of the network becoming overly specialized to a particular class of objects. In the AI-GODT case, the natural images were obtained from a publicly available Kaggle dataset [42] containing over 4,000 images of flowers. Each image was first converted to grayscale and resampled to $256 \times 256$ pixels. The resulting intensity values were subsequently normalized to the range [-1.0, 1.0]. For each sample, an amplitude modulation factor $\delta$ was randomly selected from the range (0, 0.25], and the normalized image $img$ was used to construct the object-wave amplitude according to

$$A = 1 + \delta \cdot img. \tag{A1}$$

The predefined shear vector $\boldsymbol{s}$ was then used to generate two relevant distributions: (1) the fringe modulation $A(\boldsymbol{r})A(\boldsymbol{r}+\boldsymbol{s})$, and (2) the corresponding expected amplitude input for GODT, i.e., $[A(\boldsymbol{r}) - A(\boldsymbol{r}+\boldsymbol{s})]/A(\boldsymbol{r})$. To account for subpixel displacements introduced by the shear operation, the images were shifted using cubic B-spline interpolation with interpolation order of 3.

The network architecture employs the rectified linear unit (ReLU) activation function, which is widely used in deep learning because of its computational simplicity and its ability to mitigate the vanishing-gradient problem. However, ReLU outputs zero for negative inputs, causing negative values in the target images to be truncated if predicted directly. To address this limitation in AI-GODT, a unity offset was added to the target images during training, shifting their values into the nonnegative range. After inference, the same offset was subtracted from the network outputs to recover the original image values.

The resulting simulated images formed the input–target samples for training the neural network. The complete dataset consisted of 4000 such pairs and was randomly divided into training, validation, and test subsets using an 8:1:1 ratio.

## Appendix B. Model architecture

The network, depicted in Fig. B1, is a fully convolutional residual neural network, designed to map the captured fringe modulation to the normalized amplitude finite difference used in AI-GODT. The input is a single-channel image of size 256 × 256 pixels. An initial 3 × 3 convolutional layer extracts 16 feature maps, which are subsequently processed by 12 residual blocks. Each residual block comprises two 3 × 3 convolutional layers with batch normalization and ReLU activation, together with an identity shortcut that facilitates the propagation of information through the network. A final 1 × 1 convolution projects the resulting feature maps onto a single-channel output. This output is then processed by batch normalization followed by a linear activation to produce the predicted normalized amplitude difference.
The fully convolutional architecture preserves the spatial dimensions of the input, enabling pixel-wise prediction without fully connected layers. This property allows the network to be trained on smaller images (256 × 256 pixels in this study) and subsequently applied directly to larger images, without requiring either patch-based processing and stitching or resizing the inputs to match the training dimensions.

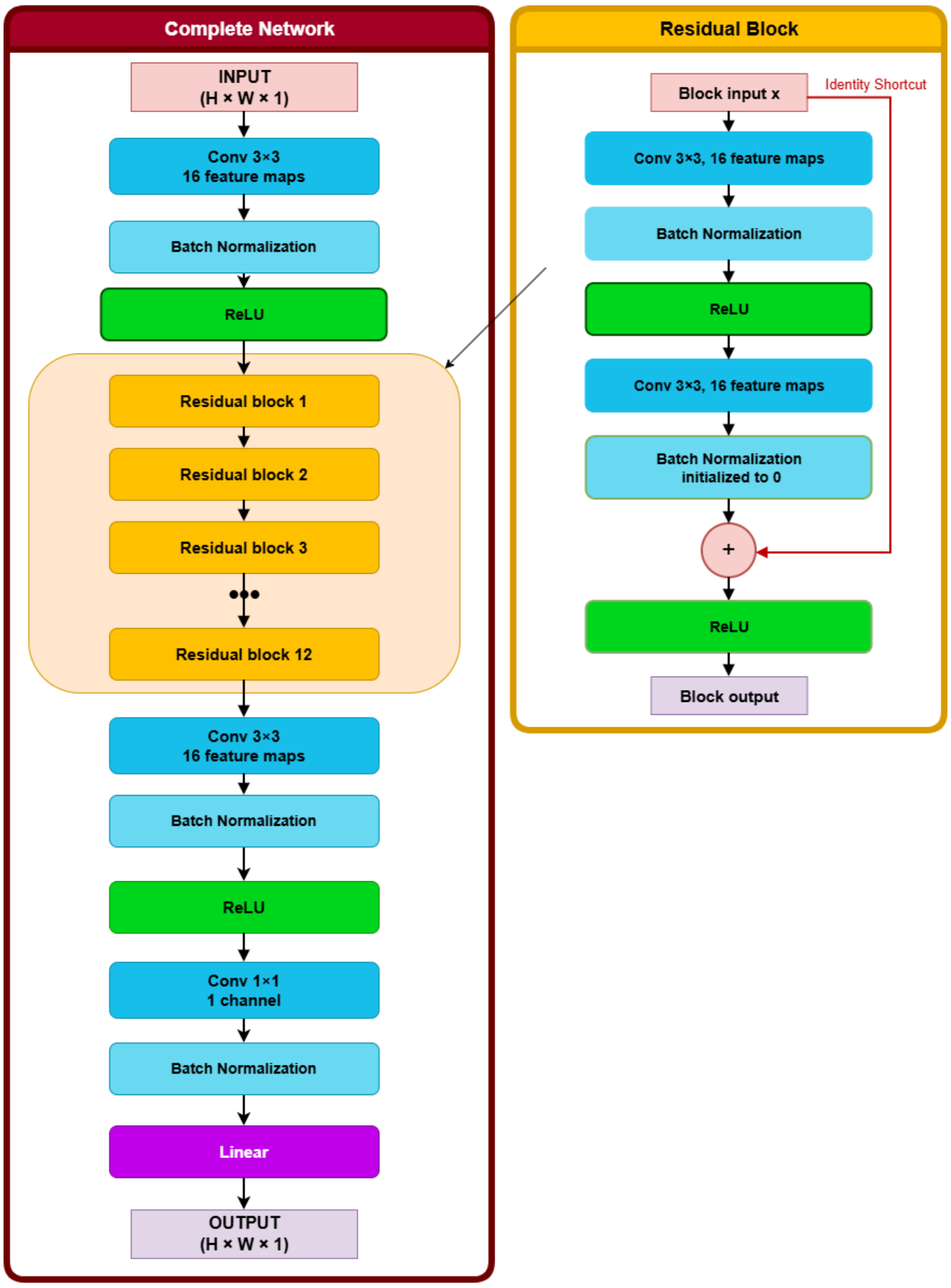

**Figure B1.** The ResNet model architecture. $H \times W \times 1$ represents the height, width, and number of channels of the input/output tensor. In this study, $H = W = 256$ during training. ReLU denotes the Rectified Linear Unit activation function, and Conv $h \times w$ represents a convolutional layer with a $h \times w$ kernel.

## Appendix C. Training

The network was trained for 25 epochs using 3,200 computer-generated input–target image pairs, with an additional 400 pairs reserved for validation after each epoch. A batch size of one was used, and the network architecture and training configuration were designed to enable rapid training on hardware with limited computational resources. The model parameters were optimized using the Adam optimizer by minimizing the mean squared error between the predicted and target normalized amplitude-difference images. An exponential learning-rate decay schedule was employed, with an initial learning rate of $5 \times 10^{-5}$ and a decay rate of 0.9 every 10,000 optimization steps.

The neural network was implemented in TensorFlow 2 and trained in a Windows Subsystem for Linux (WSL) environment. Training was completed in less than 15 minutes on a computer equipped with a 13th Gen Intel Core i9-13980HX processor (2.20 GHz), 32 GB of RAM, and an NVIDIA GeForce RTX 4080 Laptop GPU with 12 GB of graphics memory.

An exemplary loss curve, obtained for the case of the phase test target (letter G) measurement, is shown on a logarithmic scale in Fig. C1. The loss decreases rapidly during the first 10 epochs and gradually stabilizes after approximately 15 epochs. The similar trends in the training and validation losses suggest consistent convergence and no evident overfitting.

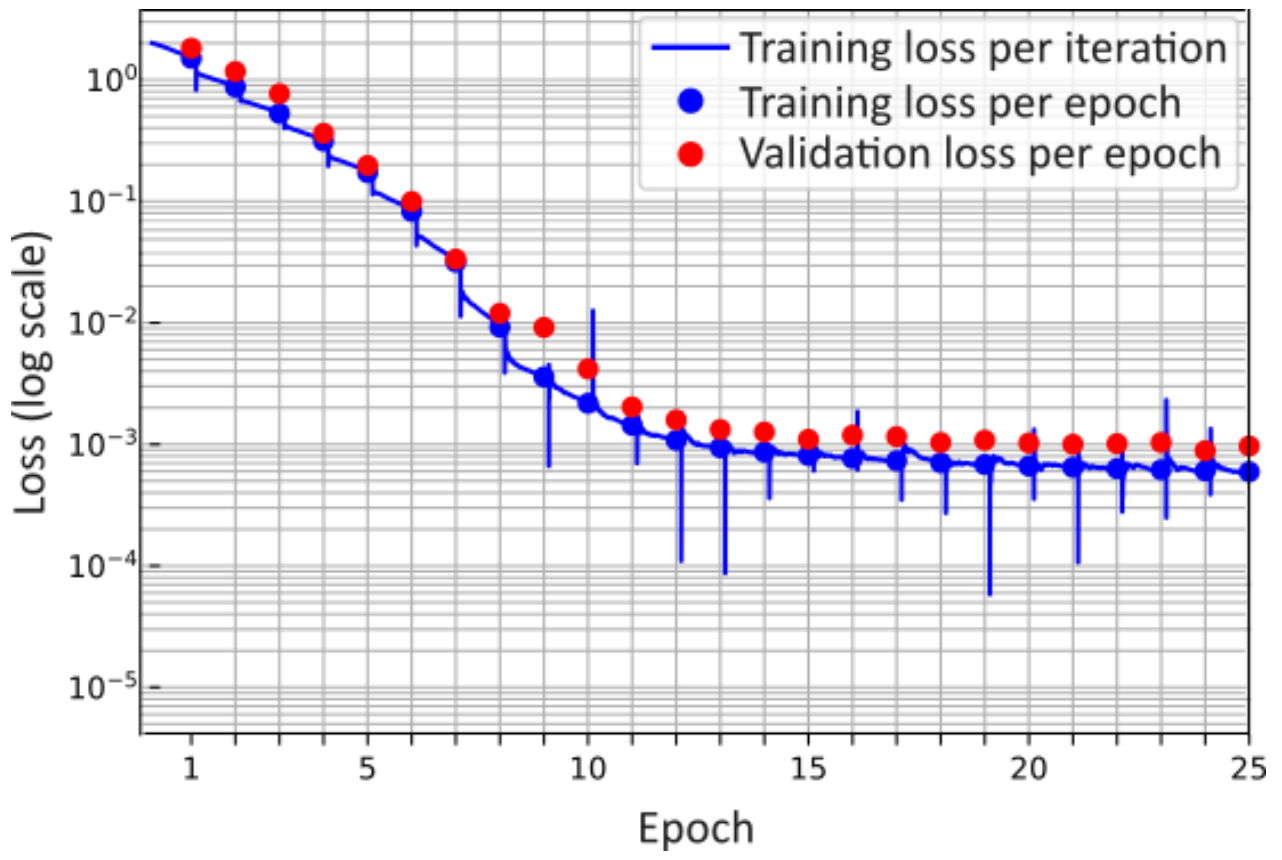

**Figure C1.** Training and validation loss curves for the ResNet model trained for the phase test target (letter G) measurement, plotted on a logarithmic scale. The blue line represents the training loss at each iteration, the blue dots indicate the training loss at the end of each epoch, and the red dots represent the validation loss for each epoch.

## Appendix D. Evaluation on test data

The trained network was evaluated on an independent test dataset generated using the same procedure as the training and validation datasets, but not used during training. Representative input, target, and predicted images are shown in Fig. D1 (a), (b), and (c), respectively. The images were obtained for $\boldsymbol{s}$ = [-4.3, 3.1] pixel. The DNN performance across the test dataset was quantified using the root mean square error (RMSE) and the structural similarity index measure (SSIM). Across all 400 test samples, the average RMSE was 0.022 ± 0.011, with a median of 0.019, while the SSIM was 0.782 ± 0.080, with a median of 0.785. For the representative example in Fig. D1 (c), the corresponding RMSE and SSIM values are also indicated to illustrate the estimation quality of individual test samples.

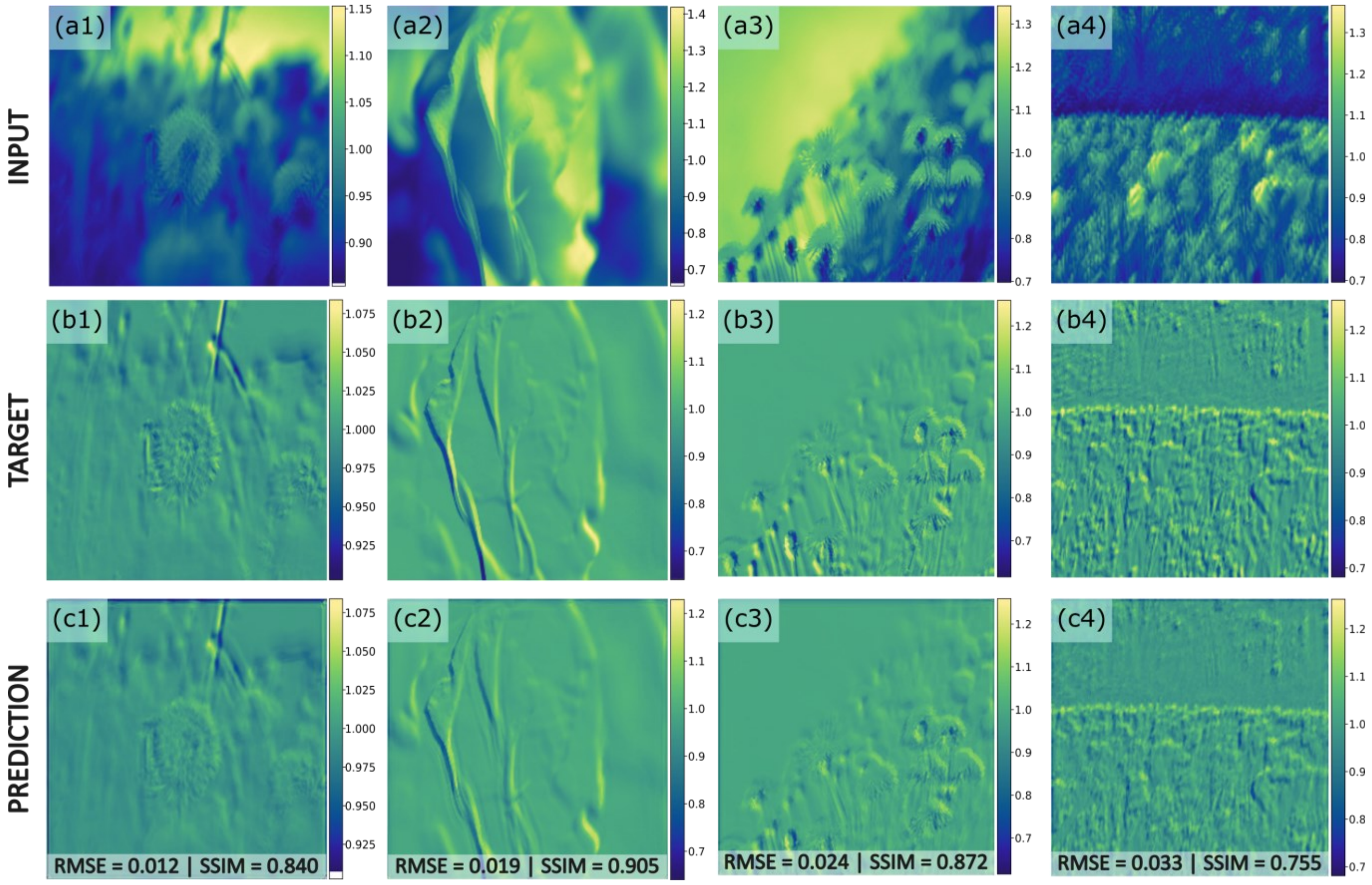

**Figure D1.** Representative results from the independent test dataset. (a1)–(a4) input images; (b1)–(b4) corresponding ground-truth target images; and (c1)–(c4) corresponding network predictions. Image values are expressed in arbitrary units. RMSE and SSIM values are indicated for each prediction to quantify the estimation accuracy.

## Appendix E. Primary hippocampal cell cultures

The neuronal cell samples presented in this study were obtained from dissociated hippocampal cultures prepared from postnatal day 0 (P0) Wistar rats [43]. Following brain extraction, the hippocampi were isolated on ice in dissociation medium (DM; 81.8 mM $Na_2SO_4$, 30 mM $K_2SO_4$, 5.8 mM $MgCl_2$, 0.25 mM $CaCl_2$, 1 mM HEPES [pH 7.4], 20 mM glucose, 1 mM kynurenic acid, and 0.001% phenol red). The isolated hippocampi were rinsed with DM and subsequently incubated twice in papain solution (100 U in DM) for 15 min at 37 °C. Following papain treatment, the tissue was rinsed three times with DM, and papain activity was terminated by incubation with a trypsin inhibitor prepared in DM.

The hippocampi were then washed three times with plating medium (PM; Minimum Essential Medium supplemented with 10% fetal bovine serum and 1% penicillin-streptomycin) and gently triturated until a cloudy cell suspension was obtained. The resulting suspension was diluted tenfold with PM and centrifuged at 1000 × g for 10 min at room temperature. The resulting cell pellets were resuspended in PM, and cell numbers were determined. Cells were subsequently plated at a density of 90,000 cells per 18-mm-diameter coverslip pre-coated with 1 mg/ml poly-D-lysine (Sigma-Aldrich) and 2.5 µg/ml laminin (Roche).

Two hours after plating, the PM was replaced with maintenance medium (MM), consisting of Neurobasal-A supplemented with 2% B-27 and 1% penicillin-streptomycin.

## References

[1] G. Kim, H. Hugonnet, K. Kim, J.-H. Lee, S.S. Lee, J. Ha, C. Lee, H. Park, K.-J. Yoon, Y. Shin, G. Csucs, I. Hitchcock, L. Mackinder, J.H. Kim, T.H. Hwang, S. Lee, P. O'Toole, B.-K. Koo, J. Guck, Y. Park, Holotomography, Nat. Rev. Methods Primer 4 (2024) 51. https://doi.org/10.1038/s43586-024-00327-1.

[2] Tomocube, Holotomography Label-Free 3D Live Cell Imaging (2026). https://www.tomocube.com/ (accessed October 29, 2026).

[3] Nanolive, Complete Solut. Your Label-Free Live Cell Imaging (2026). https://www.nanolive.com/ (accessed October 29, 2026).

[4] Y. Sung, W. Choi, C. Fang-Yen, K. Badizadegan, R.R. Dasari, M.S. Feld, Optical diffraction tomography for high resolution live cell imaging, Opt. Express 17 (2009) 266–277. https://doi.org/10.1364/OE.17.000266.
[5] M. Rogalski, J. Winnik, J. Dudek, P. Arcab, E. Wdowiak, P. Matryba, M. Stefaniuk, P. Zdańkowski, M. Trusiak, Gigavoxel-scale multiple-scattering-aware lensless holotomography, Light Sci. Appl. 15 (2026) 381. https://doi.org/10.1038/s41377-026-02416-0.
[6] J.W. Goodman, Some fundamental properties of speckle*, JOSA 66 (1976) 1145–1150. https://doi.org/10.1364/JOSA.66.001145.
[7] P. Zdańkowski, J. Winnik, K. Patorski, P. Gocłowski, M. Ziemczonok, M. Józwik, M. Kujawińska, M. Trusiak, Common-path intrinsically achromatic optical diffraction tomography, Biomed. Opt. Express 12 (2021) 4219–4234. https://doi.org/10.1364/BOE.428828.
[8] P. Zdańkowski, J. Winnik, M. Rogalski, M.J. Marzejon, E. Wdowiak, W. Dudka, M. Józwik, M. Trusiak, Polarization Gratings Aided Common-Path Hilbert Holotomography for Label-Free Lipid Droplets Content Assay, Laser Photonics Rev. 19 (2025) 2401474. https://doi.org/10.1002/lpor.202401474.
[9] Z. Wang, D.L. Marks, P.S. Carney, L.J. Millet, M.U. Gillette, A. Mihi, P.V. Braun, Z. Shen, S.G. Prasanth, G. Popescu, Spatial light interference tomography (SLIT), Opt. Express 19 (2011) 19907–19918. https://doi.org/10.1364/OE.19.019907.
[10] P. Goclowski, J. Winnik, V. Dubey, P. Zdankowski, M. Trusiak, U. Neogi, M. Varshney, B.S. Ahluwalia, A. Ahmad, Lateral shearing optical diffraction tomography of brain organoid with reduced spatial coherence, arXiv:2601.20082 (2026). https://doi.org/10.48550/arXiv.2601.20082.
[11] P. Bouchal, R. Chmelík, Z. Bouchal, Phase of white light and its compatibility to the optical path, Opt. Express 29 (2021) 12398–12412. https://doi.org/10.1364/OE.418878.
[12] W. Lipke, J. Winnik, M. Trusiak, Numerical analysis of the effect of reduced temporal coherence in quantitative phase microscopy and tomography, Opt. Express 30 (2022) 21241–21257. https://doi.org/10.1364/OE.458167.
[13] Y. Kim, H. Shim, K. Kim, H. Park, J.H. Heo, J. Yoon, C. Choi, S. Jang, Y. Park, Common-path diffraction optical tomography for investigation of three-dimensional structures and dynamics of biological cells, Opt. Express 22 (2014) 10398–10407. https://doi.org/10.1364/OE.22.010398.
[14] Z. Tang, J. Winnik, B.M. Hennelly, Optical diffraction tomography using a self-reference module, Biomed. Opt. Express 16 (2025) 57–67. https://doi.org/10.1364/BOE.545296.
[15] M. Chen, L. Tian, L. Waller, 3D differential phase contrast microscopy, Biomed. Opt. Express 7 (2016) 3940–3950. https://doi.org/10.1364/BOE.7.003940.
[16] J.M. Soto, J.A. Rodrigo, T. Alieva, Optical diffraction tomography with fully and partially coherent illumination in high numerical aperture label-free microscopy [Invited], Appl. Opt. Vol 57 Issue 1 Pp A205-A214 (2018). https://doi.org/10.1364/AO.57.00A205.
[17] C. Park, K. Lee, Y. Baek, Y. Park, Low-coherence optical diffraction tomography using a ferroelectric liquid crystal spatial light modulator, Opt. Express 28 (2020) 39649–39659. https://doi.org/10.1364/OE.405418.
[18] I. Michálková, V. Kollárová, M. Špaček, R. Chmelík, Dataset-driven holographic incoherent-light-source optical diffraction tomography applied to red blood cells, J. Phys. Photonics 8 (2026) 025048. https://doi.org/10.1088/2515-7647/ae7ad2.
[19] Y. Li, Y. Xue, R. Liu, Y. Feng, J. Liang, C. Bai, T. Wu, L. Yang, X. Yu, X. Xu, K. Wen, P. Gao, T. Dai, L. Kong, J. Min, B. Yao, Optical diffraction tomography using programmable LED array illumination and quantitative phase camera, J. Phys. Photonics 8 (2026) 035005. https://doi.org/10.1088/2515-7647/ae82d2.
[20] D.-M. Ta, M. Ziemczonok, A. Kus, P. Bon, Wavefront imaging with low coherence illumination simplifies the optical diffraction tomography setup and improves the signal-to-noise ratio, Opt. Lett. 51 (2026) 3052–3055. https://doi.org/10.1364/OL.592719.
[21] S. Bianchi, F. Brasili, F. Saglimbeni, B. Cortese, R.D. Leonardo, Optical diffraction tomography of 3D microstructures using a low coherence source, Opt. Express 30 (2022) 22321–22332. https://doi.org/10.1364/OE.454910.

[22] X. Yuan, J. Min, Y. Zhou, Y. Xue, C. Bai, M. Li, X. Xu, B. Yao, Optical diffraction tomography based on quadriwave lateral shearing interferometry, Opt. Laser Technol. 177 (2024) 111124. https://doi.org/10.1016/j.optlastec.2024.111124.

[23] J. Winnik, P. Zdańkowski, M. Stefaniuk, A. Ahmad, C. Zuo, B.S. Ahluwalia, M. Trusiak, Gradient optical diffraction tomography, Commun. Phys. 8 (2025) 510. https://doi.org/10.1038/s42005-025-02416-3.

[24] K.-N. Joo, H.M. Park, Shearing Interferometry: Recent Research Trends and Applications, Curr. Opt. Photonics 7 (2023) 325–336. https://doi.org/10.3807/COPP.2023.7.4.325.

[25] G. Baffou, Quantitative phase microscopy using quadriwave lateral shearing interferometry (QLSI): principle, terminology, algorithm and grating shadow description, J. Phys. Appl. Phys. 54 (2021) 294002. https://doi.org/10.1088/1361-6463/abfbf9.

[26] P. Gocłowski, H. Mao, M. Trusiak, B.S. Ahluwalia, A. Ahmad, High-fidelity single-shot quantitative differential phase microscopy using a pseudothermal Sagnac interferometer, Opt. Lett. 51 (2026) 4127–4130. https://doi.org/10.1364/OL.599071.

[27] J. Winnik, D. Suski, M. Heto, M. Lenarcik, M. Ziemczonok, M. Kołodziejczyk, M. Trusiak, P. Zdańkowski, OSI-flex: optimization-based shearing interferometry for joint phase and shear estimation using a flexible open-source framework, J. Phys. Photonics 8 (2026) 015036. https://doi.org/10.1088/2515-7647/ae3504.

[28] Phasics, QWLSI Powerful Wavefront Technol. Overcoming Limit. Shack-Hartmann Wavefront Sens. Fizeau Interf. (2026). https://phasics.com (accessed October 29, 2026).

[29] Y. Xue, J. Min, L. Yang, T. Wu, M. Li, C. Bai, X. Yu, X. Xu, S. Yan, B. Yao, High-resolution quantitative phase camera utilizing a polarization-modulated diffractive optical element, Optica 12 (2025) 1820–1829. https://doi.org/10.1364/OPTICA.577330.

[30] N. Ali, W. Wang, J. Winnik, Z. Xiong, R. Yan, H. Cheng, Y. Dong, P. Zdańkowski, M. Trusiak, P. Gao, Refractive index gradient tomography for defect detection, Appl. Opt. 65 (2026) 9826–9834. https://doi.org/10.1364/AO.610310.

[31] E. Wolf, Three-dimensional structure determination of semi-transparent objects from holographic data, Opt. Commun. 1 (1969) 153–156. https://doi.org/10.1016/0030-4018(69)90052-2.

[32] S. Pan, A. Kak, A computational study of reconstruction algorithms for diffraction tomography: Interpolation versus filtered-backpropagation, IEEE Trans. Acoust. Speech Signal Process. 31 (1983) 1262–1275. https://doi.org/10.1109/TASSP.1983.1164196.

[33] K. He, X. Zhang, S. Ren, J. Sun, Deep Residual Learning for Image Recognition, in: 2016 IEEE Conf. Comput. Vis. Pattern Recognit. CVPR, 2016: pp. 770–778. https://doi.org/10.1109/CVPR.2016.90.

[34] Julianna Winnik, Konrad Kosuge, QCI-LAB/AI-GODT: Amplitude-Informed Gradient Optical Diffraction Tomography, (n.d.). https://github.com/QCI-LAB/AI-GODT (accessed November 5, 2026).

[35] J. Winnik, P. Zdańkowski, K. Kosuge, A. Parys, M. Stefaniuk, M. Trusiak, AI-GODT – Amplitude-Informed Gradient Optical Diffraction Tomography Dataset, (2026). https://doi.org/10.5281/zenodo.24342497.

[36] Julianna Winnik, QCI-LAB/GRAD-EWALD: GRAD-EWALD: optical diffraction tomography reconstruction software with gradient data support, (n.d.). https://github.com/QCI-LAB/GRAD-EWALD (accessed October 30, 2026).

[37] biopto/EWALD: EWALD: optical diffraction tomography reconstruction software, (n.d.). https://github.com/biopto/EWALD (accessed October 30, 2026).

[38] M. Rogalski, P. Arcab, L. Stanaszek, V. Micó, C. Zuo, M. Trusiak, Physics-driven universal twin-image removal network for digital in-line holographic microscopy, Opt. Express 32 (2024) 742–761. https://doi.org/10.1364/OE.505440.

[39] J. Winnik, A. Walocha, W. Ogonowski, W. Forjasz, P. Arcab, M. Rogalski, A. Rutkowska, M. Stefaniuk, J.Á. Picazo-Bueno, V. Micó, M. Trusiak, M. Cywińska, YOSO: single-frame Gerchberg–Saxton phase retrieval with AI-based data augmentation for in-line holography, J. Phys. Photonics 8 (2026) 035036. https://doi.org/10.1088/2515-7647/ae99bb.

[40] K.-H. Brenner, W. Singer, Light propagation through microlenses: a new simulation method, Appl. Opt. Vol 32 Issue 26 Pp 4984-4988 (1993). https://doi.org/10.1364/AO.32.004984.

[41] J. Kostencka, T. Kozacki, A. Kuś, B. Kemper, M. Kujawińska, Holographic tomography with scanning of illumination: space-domain reconstruction for spatially invariant accuracy, Biomed. Opt. Express 7 (2016) 4086–4101. https://doi.org/10.1364/BOE.7.004086.
[42] S. Gupta, Flowers Dataset, (2020). https://www.kaggle.com/datasets/imsparsh/flowers-dataset (accessed October 5, 2026).
[43] M. Roszkowska, A. Krysiak, L. Majchrowicz, K. Nader, A. Beroun, P. Michaluk, M. Pekala, J. Jaworski, L. Kondrakiewicz, A. Puścian, E. Knapska, L. Kaczmarek, K. Kalita, SRF depletion in early life contributes to social interaction deficits in the adulthood, Cell. Mol. Life Sci. 79 (2022) 278. https://doi.org/10.1007/s00018-022-04291-5.